\documentclass[doublecol]{epl2} 
\usepackage{graphicx}% Include figure files
\usepackage{dcolumn}% Align table columns on decimal point
\usepackage{bm}% bold math
\usepackage[utf8]{inputenc}
\usepackage[T1]{fontenc}
\usepackage{mathptmx}
\usepackage{etoolbox}
\usepackage{cite}
\usepackage{color}
\usepackage{amsmath}
\usepackage{amssymb}
\usepackage{subfigure}
\usepackage{epsfig}
\usepackage{float}
\usepackage{lipsum}
\usepackage{cases}
\usepackage{appendix}
\usepackage{cancel}
\usepackage{soul}
\usepackage{pdfpages}
\newcommand{\G}{\hat{\mathrm{G}}}

\newcommand{\F}{\hat{\mathrm{F}}}

\newcommand{\T}{\hat{\mathrm{T}}}

\newcommand{\1}{\mathbb{I}}

\newcommand{\qq}{\bm{q}}

\newcommand{\dif}{\mathop{}\!\mathrm{d}}

\newcommand{\V}{\mathop{}\!\mathrm{V}}

\newcommand{\diag}{\mathop{}\!\bm{\mathrm{diag}}}

\newcommand{\vv}{\bm{v}} 

\newcommand{\rr}{\bm{r}}

\newcommand{\xxi}{\bm{\xi}}

\newcommand{\I}{\mathrm{I}}
\newcommand{\U}{\hat{\mathrm{U}}}

\newcommand{\dbar}{\lower0.15ex\hbox{$\mathchar'26$}\mkern-12mu \dif}

\newcommand{\amnt}[1]{{\color{black}{#1}}} %amendment
\title{\amnt{Escape Dynamics  in an Anisotropically Driven Brownian Magneto-System}
\shorttitle{Escape Dynamics  in an Anisotropically Driven Brownian Magneto-System}}
\author{I. Abdoli\inst{1} \and J.-U. Sommer\inst{1, 2} \and H. L\"owen\inst{3} \and A. Sharma\inst{1, 2}}
\shortauthor{I. Abdoli  \etal}
 
\institute{                    
  \inst{1} Leibniz-Institut  f\"ur Polymerforschung Dresden, Institut Theorie der Polymere, 01069 Dresden, Germany \\
  \inst{2} Technische Universit\"at Dresden, Institut f\"ur Theoretische Physik, 01069 Dresden, Germany \\
\inst{3} Institut f\"ur Theoretische Physik II: Weiche Materie, Heinrich-Heine-Universit\"at D\"usseldorf, D\"usseldorf, 40225, Germany}

\abstract{
Thermally activated escape of a Brownian particle over a potential barrier is well understood within Kramers theory. When subjected to an external magnetic field, the Lorentz force slows down the escape dynamics via a rescaling of the diffusion coefficient without affecting the exponential dependence on the barrier height. Here, we study the escape dynamics of a charged Brownian particle from a two-dimensional truncated harmonic potential under the influence of Lorentz force due to an external magnetic field. \amnt{The particle is driven anisotropically by subjecting it to noises with different strengths along different spatial directions. We show that the escape time can largely be tuned by the anisotropic driving.  While the escape process becomes anisotropic due to the two different noises, the spatial symmetry is restored in the limit of large magnetic fields. This is attributed to the Lorentz force induced coupling between the spatial degrees of freedom which makes the difference between two noises irrelevant at high magnetic fields.} The theoretical predictions are verified by Brownian dynamics simulations. In principle, our predictions can be tested by experiments with a Brownian gyrator in the presence of a magnetic field.
}

\begin{document}
\maketitle
\section{Introduction}
Escape of a particle from a metastable potential well is one of the most celebrated problems in statistical physics~\cite{hanggi1990reaction}. The escape problem was first conceived and studied by Kramers to model the rate of chemical reactions~\cite{kramers1940brownian}. In the classical escape problem, which includes only  thermal fluctuations, the escape rate decreases exponentially with the barrier height.

The classical trap model has been generalized to include fluctuations other than thermal fluctuations~\cite{hanggi1984bistable,jung1988bistability, sharma2017escape,scacchi2019escape,caprini2019active}. Such a scenario arises naturally in active gels~\cite{mizuno2007nonequilibrium, sheinman2015anomalous}, in which embedded tracer particles are subjected to both thermal fluctuations and motor-induced, athermal fluctuations~\cite{woillez2020active}, and for active particles~\cite{burada2012escape, koumakis2014directed, schneider2019optimal, wexler2020dynamics, debnath2021escape, schwarzendahl2021barrier, zanovello2021optimal}. While one models the thermal fluctuations in the usual fashion as white Gaussian noise, the other noise is often described as a colored noise to model the effect of active, athermal fluctuations. The presence of an additional noise can dramatically modify the escape dynamics of a particle~\cite{woillez2020active,militaru2021escape, caprini2021correlated, gera2021solution}. 

%\amnt{Here we explore the escape dynamics of a single Brownian particle from a potential well subjected to thermal noises with different strengths along different spatial directions, i.e., acting on the different Cartesian components of the momentum vector. Such a system constitutes a primitive Brownian engine composed of a gyrating Brownian particle~\cite{filliger2007brownian}. This kind of anisotropic driving can be experimentally realized by applying a strongly fluctuating electric field to a charged Brownian particle in one direction mimicking the role of an additional temperature~\cite{argun2017experimental}. Consequently, the effective temperature along this direction takes a value which is higher than the temperature of the bath.  We show that breaking the spatial symmetry via two different noises results in anisotropic escape dynamics; more particles escape the potential well along the axis with larger noise strength which we refer to as the hot axis. This gives rise to a large tunability of the escape time by the difference between noise strengths or effective temperature difference. However, when the particle is further subjected to Lorentz force due to an external magnetic field, we show that the spatial symmetry is restored in the limit of large magnetic fields; the escape is driven by an effective noise with the average strength of the two noises. This is attributed to the Lorentz force induced coupling between the spatial degrees of freedom which makes the difference between noise strengths irrelevant at high magnetic fields. }
\amnt{Here we explore the escape dynamics of a single Brownian particle from a potential well subjected to two noises with different strengths along different spatial directions, i.e., acting on the different Cartesian components of the momentum vector. Such a system constitutes a primitive Brownian engine composed of a gyrating Brownian particle~\cite{filliger2007brownian}. This kind of anisotropic driving can be experimentally realized by applying a strongly fluctuating electric field to a charged Brownian particle in one direction mimicking the role of an additional temperature~\cite{argun2017experimental}. Consequently, the effective temperature along this direction takes a value which is higher than the temperature of the bath. 
We show that breaking the spatial symmetry via two different noises results in anisotropic escape dynamics; more particles escape the potential well along the axis with larger noise strength which we refer to as the hot axis. This gives rise to a large tunability of the escape time by the difference between noise strengths or effective temperature difference. However, when the particle is further subjected to Lorentz force due to an external magnetic field, we show that the spatial symmetry is restored in the limit of large magnetic fields; the escape is driven by an effective noise with the average strength of the two noises. This is attributed to the Lorentz force induced coupling between the spatial degrees of freedom which makes the difference between noise strengths irrelevant at high magnetic fields. }

%\amnt{In a spatially isotropic Brownian magneto-system, the Lorentz force slows down the escape dynamics via a trivial rescaling of the diffusion coefficient~\cite{filliger2007kramers, abdoli2020nondiffusive, sm} without affecting the exponential dependence on the barrier height. In contrast, in an anisotropically driven system, a magnetic field affects the dynamics in a qualitatively different way~\cite{abdoli2020correlations}. While curving the trajectory of a particle, there also occurs energy transfer in form of heat from the hot source with a higher effective temperature to the cold source with lower effective temperature, which is mediated by the magnetic field~\cite{abdoli2021brownian}. As a consequence, at large magnetic fields, the two spatial degrees of freedom become identical regardless of the effective temperature difference. In general, it is difficult to study the escape problem in a multidimensional potential well. Here we use an asymptotic method to compute the mean escape time for a large barrier height, obtaining an explicit analytical prediction.}
\amnt{In a spatially isotropic Brownian magneto-system, the Lorentz force slows down the escape dynamics via a trivial rescaling of the diffusion coefficient~\cite{filliger2007kramers, abdoli2020nondiffusive, sm} without affecting the exponential dependence on the barrier height. In contrast, in an anisotropically driven system, a magnetic field affects the dynamics in a qualitatively different way~\cite{abdoli2020correlations}. While curving the trajectory of a particle, there also occurs energy transfer in form of heat from the hot source with a higher effective temperature to the cold source with lower effective temperature, which is mediated by the magnetic field~\cite{abdoli2021brownian}. As a consequence, at large magnetic fields, the two spatial degrees of freedom become identical regardless of the effective temperature difference. In general, it is difficult to study the escape problem in a multidimensional potential well. Here we use an asymptotic method to compute the mean escape time for a large barrier height, obtaining an explicit analytical prediction.}

\section{\label{model}\amnt{Anisotropically driven Brownian magneto-system}}
\begin{figure}[t]
\centering
\includegraphics[width=0.6\linewidth]{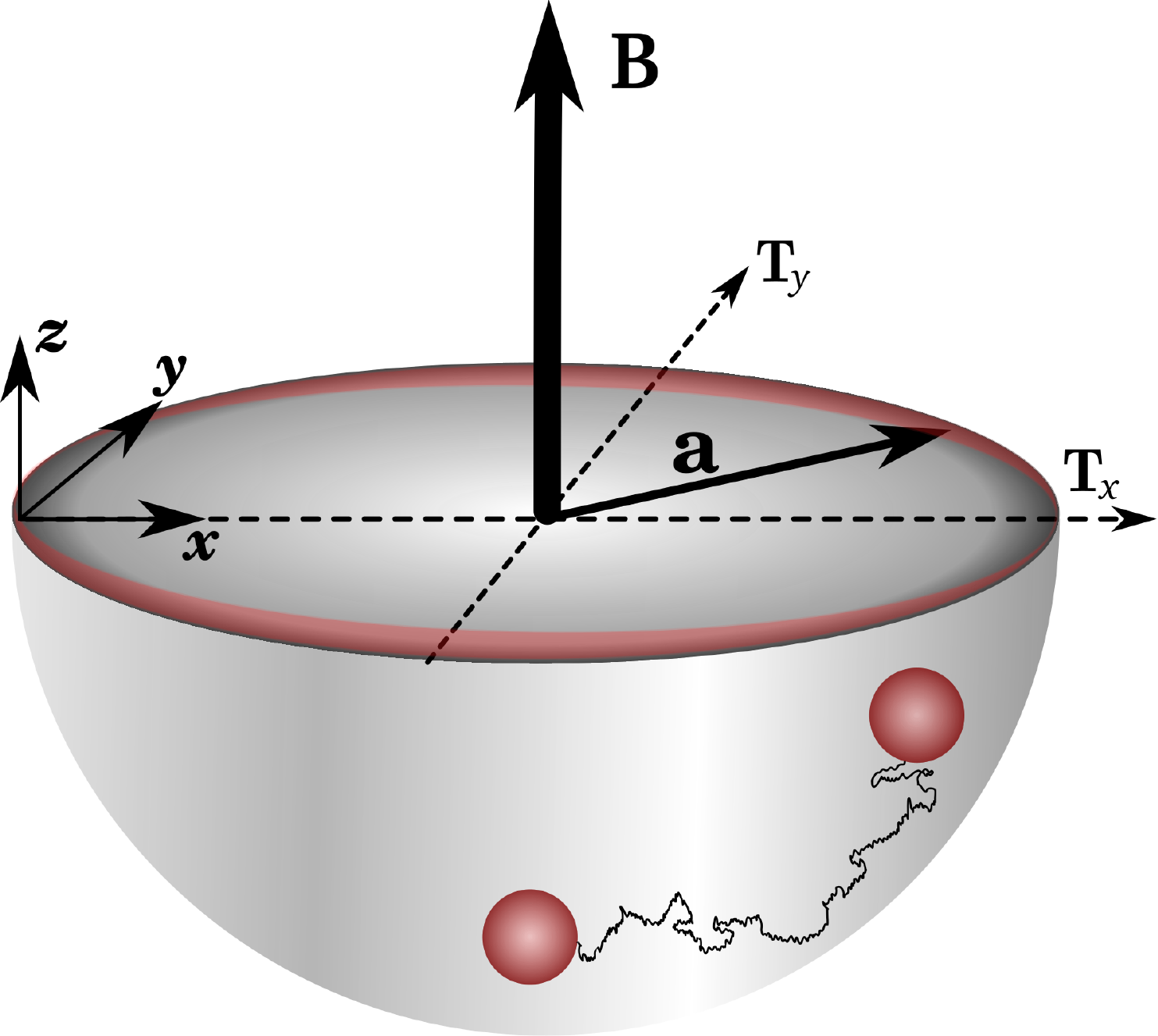}
\caption{A single charged Brownian particle, subjected to an external magnetic field $B$ and trapped in an isotropic harmonic potential $V(x,y)=k(x^2+y^2)/2$ with the parameter $k$. The particle is simultaneously subjected to \amnt{two noises with different strengths along its $x$ and $y$ degrees of freedom. The particle escapes more along the $y$ axis due to the broken spatial symmetry by the two noises, which is shown in reddish color.} We show that escape distribution can be tuned and symmetrized by varying the magnitude of the applied magnetic field. 
%A single charged Brownian particle, subjected to an external magnetic field $B$ and trapped in an isotropic harmonic potential $V(x,y)=k(x^2+y^2)/2$ with the parameter $k$. The particle is simultaneously subjected to two different thermal noises coupled to its $x$ and $y$ degrees of freedom. The particle escapes more along the $y$ axis due to the broken spatial symmetry by the two thermostats, which is shown in reddish color. We show that escape distribution can be tuned and symmetrized by varying the magnitude of the applied magnetic field.
}

\label{fig1:schema}
\end{figure} 
\amnt{We consider a Brownian magneto-system, which is made of a single diffusing particle of the charge $q$ and mass $m$ in the presence of a constant magnetic field $B$ in the $z$ direction. The $x$ and $y$ positions of the particle, which we indicate in the vectorial form $\rr=(x, y)^\top$, are coupled to noises with different strengths mapping to effective temperatures $T_x$ and $T_y$, respectively (see Fig.~\ref{fig1:schema}). The transpose is indicated by $\top$. Since the Lorentz force does not affect the motion of the particle in the direction of the applied magnetic field, we effectively investigate a two-dimensional system in the $xy$ plane. }
%We consider a two-temperature Brownian magneto-system, which is made of a single diffusing particle of the charge $q$ and mass $m$ in the presence of a constant magnetic field $B$ in the $z$ direction. The $x$ and $y$ positions of the particle, which we indicate in the vectorial form $\rr=(x, y)^\top$, are coupled to thermostats at temperatures $T_x$ and $T_y$, respectively (see Fig.~\ref{fig1:schema}). The transpose is indicated by $\top$. Since the Lorentz force does not affect the motion of the particle in the direction of the applied magnetic field, we effectively investigate a two-dimensional system in the $xy$ plane. 

The dynamics of the particle with velocity $\vv=(v_x, v_y)^\top$ which is trapped in a potential in the form of $\V(\rr)=\frac{1}{2}\rr^\top\cdot\U\cdot\rr$  can be  described by the following underdamped Langevin equation:

\begin{equation}
m\dot{\qq}(t) =  -\F\qq(t) + \xxi(t),
\label{langevinequation:q}
\end{equation}
where  $\qq(t)=(x(t), y(t), v_x(t), v_y(t))^\top$ and $\xxi(t)=(0, 0, \xi_x(t),\xi_y(t))^\top$ is Gaussian white noise with zero mean and time correlation $\langle\xxi(t)\xxi^{\top}(t')\rangle= 2\gamma\T\delta(t-t')$ where $\gamma$ is the friction coefficient. Throughout the paper we set the Boltzmann constant $k_B$ to unity. \amnt{For an isotropic potential $\U=k\hat{\1}$ with $k$ being the stiffness of the potential.}
We consider a reference system at temperature $T$. The two \amnt{effective} temperatures along Cartesian coordinates are measured in units of $T$. Length and time are measured in units of $\sqrt{T/k}$ and $\gamma/k$, respectively. 
Here $\T=\diag(0, 0, T_x, T_y)$ is a diagonal  matrix and the matrix $\F$ is defined as
\begin{equation}
\label{methods:matrixF}
\F = \left( \begin{array}{cc}
\hat{\mathrm{0}} & -m\hat{\1} \\
\U & \G \\
\end{array}\right), \,\,\, \text{where} \,\,\,\, \G = \gamma\left( \begin{array}{cc}
1 & -\kappa \\
\kappa & 1 \\
\end{array}\right).
\end{equation}
Here $\hat{\1}$ is the identity matrix and $\kappa=qB/\gamma$ is the diffusive Hall parameter which quantifies the strength of the Lorentz force relative to the frictional force. Note that the hat over the symbols indicates the \amnt{matrices} and the vectors are shown by bold symbols.

\amnt{We are interested in the overdamped dynamics of the particle. Usually this is done by setting the inertia term in Eq.~\eqref{langevinequation:q} to zero. However, in the presence of a magnetic field, this yields an incorrect description of the overdamped dynamics~\cite{chun2018emergence, vuijk2019anomalous} . A careful small-mass limit of the Langevin equation reveals that the overdamped dynamics, though diffusive in nature, are described an unusual odd-diffusion tensor that has both even (diagonal) and odd (off-diagonal) elements. We note that the variance of the particle position is determined only by the even part of the diffusion tensor~\cite{abdoli2020nondiffusive, abdoli2020correlations}.  The odd part of the tensor, which has its origin in the broken time reversal symmetry due to the magnetic field, gives rise to additional Lorentz (rotational) fluxes which are missed if one takes the route of setting the inertial term to zero.  The Lorentz fluxes can play a crucial role in the dynamical properties of Brownian systems~\cite{abdoli2020nondiffusive, abdoli2020correlations}. }

\section{Escape from an isotropic trap}
We consider a trapped particle in an isotropic potential $\V(\rr)$ and postulate that the particle escapes the trap when it reaches the boundary, truncated at $r=a$, where $r=|\rr|$ is the distance from the origin, as shown in Fig.~\ref{fig1:schema}.
\amnt{An exact calculation of the mean escape time is possible in a single temperature magneto-system, as derived in details in the Supplemental Information (SI) given in~\cite{sm}, taking advantage of the spatial isotropy. However, in an anisotropically driven magneto-system, the broken spatial symmetry prevents such an approach.} We therefore use an asymptotic method to investigate the escape dynamics in our Brownian magneto-system.  
We assume that the barrier height $\Delta E=\frac{1}{2}ka^2$, is sufficiently large that the particle leaks out slowly across the trap and settles into a quasistationary state: the escape is a Poisson process with the inverse rate of mean escape time $\langle t_{esc}\rangle$. The quasistationary probability density is given by $P(\rr, t)\sim\rho_{ss}(\rr)e^{-t/\langle t_{esc}\rangle}$ where $\rho_{ss}(\rr)$ is the steady-state probability density obtained in the limit of $a\rightarrow \infty$ which we have obtained in previous works~\cite{abdoli2020correlations, abdoli2021brownian} and in the SI in~\cite{sm}.

%We consider a trapped particle in an isotropic potential $\V(\rr)$ and postulate that the particle escapes the trap when it reaches the boundary, truncated at $r=a$, where $r=|\rr|$ is the distance from the origin, as shown in Fig.~\ref{fig1:schema}. An exact calculation of the mean escape time is possible in a system with a single thermostat , as derived in details in the Supplemental Information (SI) given in~\cite{sm}, taking advantage of the spatial isotropy. However, in a two-temperature system, the broken spatial symmetry prevents such an approach. We therefore use an asymptotic method to investigate the escape dynamics in our Brownian magneto-system.  We assume that the barrier height $\Delta E=\frac{1}{2}ka^2$, is sufficiently large that the particle leaks out slowly across the trap and settles into a quasistationary state: the escape is a Poisson process with the inverse rate of mean escape time $\langle t_{esc}\rangle$. The quasistationary probability density is given by $P(\rr, t)\sim\rho_{ss}(\rr)e^{-t/\langle t_{esc}\rangle}$ where $\rho_{ss}(\rr)$ is the steady-state probability density obtained in the limit of $a\rightarrow \infty$ which we have obtained in previous works~\cite{abdoli2020correlations, abdoli2021brownian} and in the SI in~\cite{sm}.

The total outgoing flux at $r=a$  is given by 
$J(t) = -\frac{\dif}{\dif t}\int_0^{2\pi}\int_{0}^a P(r, \theta, t)r\dif r\dif\theta$ where $P(r, \theta, t)$ is the quasistationary probability distribution in the polar coordinates. The outgoing flux can be alternatively defined as the probability $a\int_0^{2\pi}\rho_{ss}(a, \theta)e^{-t/\langle t_{esc}\rangle}\dif\theta$ to be on the boundary at time $t$, times the velocity of the fluctuation path leading to the boundary~\cite{risken1996fokker,woillez2020active}. Using the two equivalent definitions of the outgoing flux the mean escape time is equal to the inverse of the stationary-state probability density on the boundary besides a prefactor. The prefactor can be determined from the exact analytical result for \amnt{a spatially isotropic magneto-system}, as shown in the SI in~\cite{sm}. By doing so, we obtain the mean escape time (see the SI in \cite{sm} for details), which can be written as 

\begin{equation}
\label{MET}
\langle t_{esc}\rangle^{-1} \approx \frac{2}{1+\kappa^2}\sqrt{\frac{T\delta_E^2}{\bar{T}(1-\delta_{\kappa}^2)}}e^{-\frac{\delta_E}{1-\delta_{\kappa}^2}}\I_0\left(-\frac{\delta_{\kappa}\delta_E}{1-\delta_{\kappa}^2}\right),
\end{equation}
where $\I_0(x)$ is the modified Bessel function of the first kind of the order zero, $\bar{T}=(T_x+T_y)/2$ is the \amnt{effective} temperature average, $\delta_E=\Delta E/\bar{T}$ is the scaled barrier height, and 
\begin{equation}
\label{deltakappa}
\delta_{\kappa}=\frac{\Delta T}{2\bar{T}\sqrt{1+\kappa^2}},
\end{equation}
is the scaled \amnt{effective} temperature difference, given as $\Delta T=T_y-T_x$, which is an explicit function of the diffusive Hall parameter.

\begin{figure}[t]
\centering
\includegraphics[width=1\linewidth]{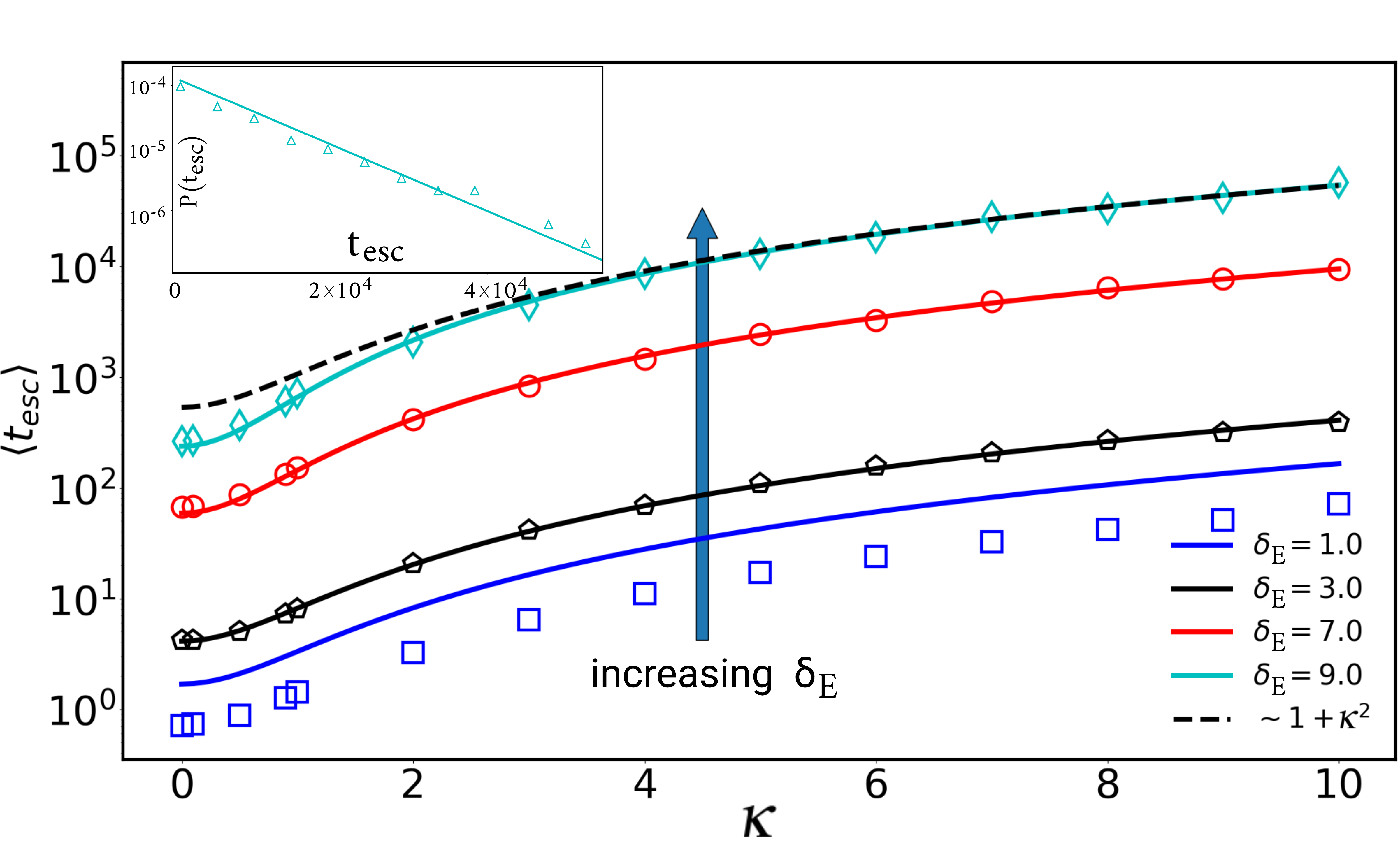}
\caption{The mean escape time as a function $\kappa$ for different barrier heights $\delta_E$ with the \amnt{effective} temperature difference $\Delta T=2.0$ and \amnt{effective} temperature average $\bar{T} = 3.0$. Solid lines show the theoretical prediction from Eq.~\eqref{MET} and the symbols depict the simulation results.  \amnt{The black, dashed line is a plot of the prefactor $1+\kappa^2$. In the limit of large magnetic fields, the overall trend of the mean escape time is determined by this prefactor.  
The inset shows the distribution of the escape time for $\kappa=4.0$ and $\delta_E=9.0$. The solid line is the Poisson distribution  $P(t)=\langle t_{esc}\rangle^{-1}\exp(-t/\langle t_{esc}\rangle)$ and the symbols depict the simulation results. The theoretical predictions overestimate the escape time for small barrier heights.}
%The mean escape time increases with increasing  magnitude of the applied magnetic field. The arrow depicts the increase in the scaled barrier height. %Our theoretical predictions are in good agreement with the simulation results.
}
  %}
\label{fig2:MTE}
\end{figure} 

\begin{figure}[t]
\centering
\includegraphics[width=1.\linewidth]{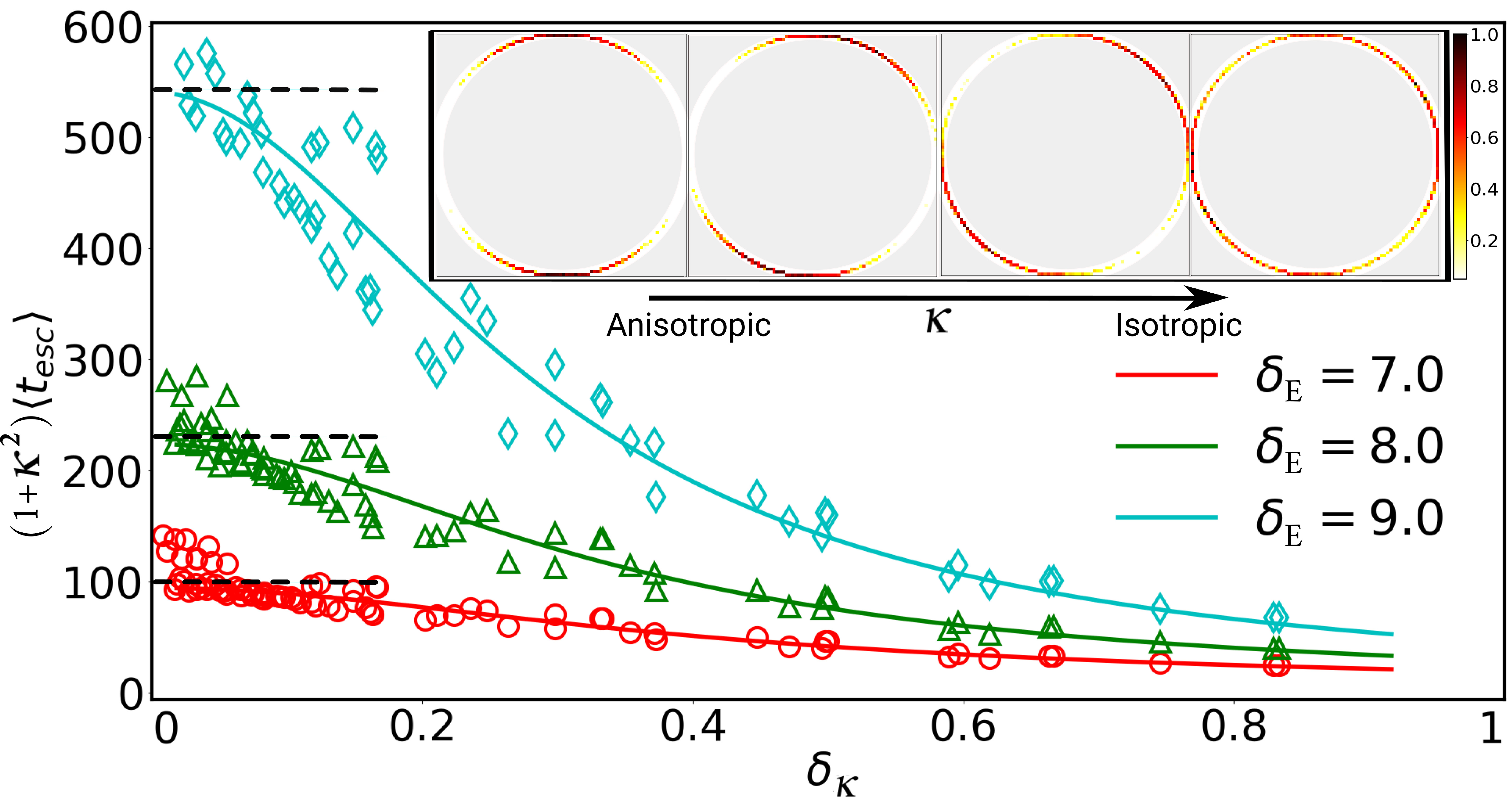}
\caption{The mean escape time as function of the scaled \amnt{effective} temperature difference $\delta_\kappa$ for different barrier heights. Solid lines show the results from the theoretical prediction in Eq.~\eqref{MET} and the symbols depict the simulation results. While a small $\delta_\kappa$ corresponds to either large magnetic fields or small \amnt{effective} temperature difference, $\delta_\kappa\rightarrow 1$ corresponds to either small magnetic fields or large \amnt{effective} temperature difference.  Simulation data is obtained for $\Delta T = 1.0, 2.0, 3.0, 4.0, 5.0$ with $\bar{T}=3.0$. The diffusive Hall parameter varies from $0.0$ to $10.0$.  
The dashed lines show the saturation to the system with temperature $\bar{T}$ corresponding to either the \amnt{effectively} zero-temperature difference  or large-magnetic-field limit. 
The inset shows the angular distribution for different values of the diffusive Hall parameter, $\kappa = 0.0, 0.5, 2.0, 10.0$ from left to right, respectively. The \amnt{effective} temperature difference is $\Delta T=2.0$ and the \amnt{effective} temperature average is $\bar{T}=3.0$. In the absence of a magnetic field the particle mostly escapes along the $y$ axis due to the broken spatial symmetry via the two \amnt{noises with different strengths, mapped to effective temperatures}. 
}
\label{fig3:comparison}
\end{figure} 
\amnt{
In a spatially isotropic system the mean escape time in Eq.~\eqref{MET} reduces to the familiar expression from Kramers theory $\langle t_{esc}\rangle^{-1}\approx (1+\kappa^2)^{-1}2\delta_E\exp(-\delta_E)$, as shown in the SI in~\cite{sm}. In the general case of two different noises, the escape time also depends on the effective temperature difference via the parameter $\delta_\kappa$. For a fixed $\kappa$, increasing the effective temperature difference accentuates the spatial anisotropy of the escape dynamics. However, in the limit of large magnetic fields $\delta_\kappa \rightarrow 0$, independent of the effective temperature difference, in which case one again obtains the Kramers escape rate for a spatially isotropic system with temperature $\bar{T}$. The $\kappa$-governed crossover from anisotropic to isotropic escape can be qualitatively understood as following. Lorentz force curves the trajectory of a moving particle. However, since the two spatial degrees of freedom are subjected to noises with different strengths, the curving of the trajectory also results in energy transfer in the form of heat from the hot source to the cold one. Consequently, in the limit of large magnetic fields, the two spatial degrees of freedom become identical regardless of the effective temperature difference.
}
%In the case of a single thermostat, the mean escape time in Eq.~\eqref{MET} reduces to the familiar expression from Kramers theory $\langle t_{esc}\rangle^{-1}\approx (1+\kappa^2)^{-1}2\delta_E\exp(-\delta_E)$, as shown in the SI in~\cite{sm}. In the general case of two different temperatures, the escape time also depends on the temperature difference via the parameter $\delta_\kappa$. For a fixed $\kappa$, increasing the temperature difference accentuates the spatial anisotropy of the escape dynamics. However, in the limit of large magnetic fields $\delta_\kappa \rightarrow 0$, independent of the temperature difference, in which case one again obtains the Kramers escape rate for a spatially isotropic system with temperature $\bar{T}$. The $\kappa$-governed crossover from anisotropic to isotropic escape can be qualitatively understood as following. Lorentz force curves the trajectory of a moving particle. However, since the two spatial degrees of freedom are coupled to different heat baths, the curving of the trajectory also results in heat flow from the hot to the cold bath. Consequently, in the limit of large magnetic fields, the two spatial degrees of freedom become identical regardless of the temperature difference.

To validate our theoretical predictions we perform Brownian dynamics simulations using the Langevin equations of motion. It has been shown that the overdamped Langevin equation for  Brownian motion in the presence of a magnetic field can give rise to unphysical values for velocity-dependent variables such as fluxes. Therefore, we use the underdamped Langevin equation, given in Eq.~\eqref{langevinequation:q}, with a sufficiently small mass. 
\amnt{ The change in the velocity due to the Gaussian noise is of the order $\sqrt{dt}/m$. By choosing the integration time step $dt < m^2$ we ensure that the velocity increments are small. Throughout the paper, we fix the mass to $m=0.001$ and the integration time step to $dt=5 \times 10^{-7}\tau$, where $\tau$ serves as a natural time scale which we set to unity.  The temperature $T$ of the reference system and the unit of the length are set to unity in the simulations. We consider a total number of particles of $1000$. }

%Throughout the paper, we fix the mass to $m=0.001$ and the integration time step to $dt=5 \times 10^{-7}\tau$, where $\tau=\gamma/k$ serves as a natural time scale which we set to unity. The temperature $T$ of the reference system and the unit of the length $\sqrt{T/k}$ serve as natural energy and length scales and are set to unity in the simulations. We consider a total number of particles of $1000$. 

Figure~\ref{fig2:MTE} shows the mean escape time as a function of the diffusive Hall parameter $\kappa$ for different values of the scaled barrier height $\delta_E$. The data are obtained from simulations of a system with the \amnt{effective} temperature average $\bar{T}=3.0$ and  \amnt{effective} temperature difference $\Delta T=2.0$. Our theoretical predictions are in good agreement with the simulation results. \amnt{In the SI given in~\cite{sm} we show that the large barrier approximation of the exact expression for escape time in a single temperature system is highly accurate compared to the known Kramers results when the expression is scaled by an empirical factor $1.2$. We use the same factor in our asymptotic approach and observe that our theoretical predictions are in good agreement with the simulation results. }

%Note that to compare with simulations, the theoretical predictions are scaled by a numerical factor of $1.2$. This factor has its origin in the high barrier limit approximation of the exact expression for escape time in a single temperature system (see the SI given in~\cite{sm}). \amnt{In the inset we show that the distribution of the escape time is Poissonian.}

\amnt{While in the limit of large magnetic fields the overall trend of the curves are determined by the prefactor  $1+\kappa^2$, as shown in Fig.~\ref{fig2:MTE} by dashed line, it deviates from this prefactor for small magnetic fields. In other words, while the anisotropic driving of the system effects the escape dynamics for small magnetic fields, the spatial symmetry is restored in the limit of large magnetic fields. }

Figure~\ref{fig3:comparison} shows the scaled mean escape time as a function of the parameter $\delta_\kappa$. The Brownian dynamics simulation has been done for systems with  \amnt{effective} temperature differences varying from $\Delta T = 1.0$ to $5.0$ and the same  \amnt{effective} temperature average $\bar{T}=3.0$. The diffusive Hall parameter is varying from $\kappa=0.0$ to $10.0$.
In the limit of large magnetic fields or small temperatures, the scaled  \amnt{effective} temperature difference parameter $\delta_\kappa \rightarrow 0$. Small magnetic fields or large  \amnt{effective} temperature difference correspond to $\delta_\kappa \rightarrow 1$. The scaled mean escape time decreases with increasing parameter $\delta_\kappa$. The dashed lines show the saturation to a system with temperature $\bar{T}$ corresponding to  \amnt{effectively} zero-temperature difference or alternatively to large-magnetic-field limit.  In the inset, we show the angular distribution from simulations for different values of the diffusive Hall parameter. %For this system, the temperature difference is $\Delta T=2.0$ and the temperature average $\bar{T}=3.0$. 
As can be seen, in the absence of a magnetic field the particle mostly escapes along the $y$ axis due to the broken spatial symmetry via \amnt{the two different noises}. The escape becomes symmetric in the limit of a large magnetic field, namely the  \amnt{effective} temperature difference becomes irrelevant.

\section{\label{conclusions}Conclusions}

\amnt{We studied the escape dynamics in an anisotropically driven Brownian magneto-system. We derived an approximate expression for the mean escape time taking into account two noises with different strength along different spatial directions, mapped to two effective temperatures. We showed that Lorentz force induces a coupling between the spatial degrees of freedom which, in the limit of large magnetic fields, restores the spatial symmetry; the two spatial degrees of freedom become identical regardless of the \amnt{effective} temperature difference.}
%We studied the escape dynamics in a two-temperature Brownian magneto-system. We derived an approximate expression for the mean escape time taking into account the average and difference of the two temperatures. We showed that Lorentz force induces a coupling between the spatial degrees of freedom which, in the limit of large magnetic fields, restores the spatial symmetry; the two spatial degrees of freedom become identical regardless of the temperature difference.

%The effect of the Lorentz force on colloidal particles can be experimentally observed when it is comparable to the frictional force, namely when the diffusive Hall parameter is comparable to unity.  A potential experimental realization could be in a complex plasma where the induced Coriolis force by rotating the reference frame acts as the Lorentz force due to an external magnetic field~\cite{kahlert2012magnetizing}. Therefore, to test our model, we suggest to trap the particle in such a set-up using optical tweezers under the effect of a fluctuating electric field, mimicking the role of an additional temperature\amnt{, mapping to a higher effective temperature}. 

\amnt{
While experimental realisation of the proposed magneto-system might be particularly difficult in colloidal systems, it is more realistic in a complex plasma where extremely large magnetic fields can be generated~\cite{kahlert2012magnetizing}. In such a set up, not only can one study the escape dynamics, one can also measure the momenta governed energy flow between the hot source and the cold bath. Such heat currents have been predicted theoretically, however, the studies employed toy models in which a particle is coupled to multiple thermostats~\cite{murashita2016overdamped, lee2020exactly}. A complex plasma, with anisotropic driving, thus presents a potentially realizable system to study momenta governed heat currents. The anisotropic noise strength can also be realized by putting the charged object in contact with active particles that provide the noise.  If the latter are propelling in an anisotropic way, as for instance on an anisotropic patterned substrate~\cite{volpe2011microswimmers} or in a nematic liquid crystalline cell~\cite{zhou2017dynamic}, their effect on the charged particle would realize an anisotropic noise strength.} Other possible realizations in which a confined particle experiences random kicks in an anisotropic environment under the additional action of a Lorentz-, Coriolis- or Magnus force are also conceivable.

In future work we aim to extend our model to the escape through a narrow hole ~\cite{schuss2007narrow,bressloff2014stochastic,nayak2020escape}. An interesting generalization would be to replace the isotropic potential by an anisotropic potential where the stationary state of the system is characterized by not only a non-Boltzmann probability distribution, but also additional Lorentz fluxes~\cite{abdoli2021brownian}.  
Also, it could be interesting to study the escape in  \amnt{an anisotropically driven} Brownian magneto-system from a bistable state, which might be realized by trapping the particle using two optical tweezers~\cite{ciliberto2017experiments}. Another study of interest would be a system in the presence of a fluctuating magnetic field where a new turnover is observed as a generic signature of the system~\cite{baura2013tuning, mondal2018generic}.    
Finally, our analysis might be applicable to other systems such as chiral colloidal microswimmers in parabolic potentials~\cite{jahanshahi2017brownian}, active Janus particles in a complex plasma~\cite{nosenko2020active}, and even rotating skyrmions~\cite{zhang2018manipulation, brown2018effect}.

\section{Acknowledgments}
A. Sharma acknowledges the support by the Deutsche Forschungsgemeinschaft (DFG) within the project SH 1275/3-1.
%\bibliographystyle{eplbib}
%\bibliography{references}
%\iffalse
\providecommand{\noopsort}[1]{}\providecommand{\singleletter}[1]{#1}%

%\fi

\end{document}

\maketitle
Here we first present the stationary-state solution to the generalized Fokker-Planck equation corresponding to the Langevin equations (1) in the main text. 
Using the stationary-state probability density, we then calculate the mean escape time for the particle to escape over a two-dimensional potential well. 

\section{Stationary-State Probability Density}

We consider a charged Brownian particle of charge $q$ under the influence of \amnt{two  noises with different strength} acting on two spatial particle coordinates and in an external magnetic field $B$ in the $z$ direction. The particle is trapped in an isotropic potential $V(r)=\frac{1}{2}\rr^\top\cdot\U\cdot\rr$ where $\rr=(x, y)^\top$ is the position of the particle and $\U=k\1$ with $k$ being the stiffness of the potential. 

The generalized Fokker-Planck equation for the probability density $\rho(\rr, t)$ of finding the particle at position $\rr$ at time $t$, corresponding to the small-mass limit of the Langevin equation~(1) in the main text, can be written as
\begin{equation}
\label{FPE}
	\frac{\partial \rho(\rr, t)}{\partial t} = \nabla\cdot\left[\D\nabla \rho(\rr, t)+ k(\G^{-1}\cdot\rr)\rho(\rr, t)\right],
\end{equation}
where 
\begin{equation}
\label{methods:matrixF}
\G = \gamma\left( \begin{array}{cc}
1 & -\kappa \\
\kappa & 1 \\
\end{array}\right),
\end{equation}
and the matrix $\D$ is given as~\cite{abdoli2020correlations,abdoli2021brownian}

\begin{equation}
\D
= \frac{1}{\gamma} 
\left( \begin{array}{cc}
\frac{T_x + \kappa^2 T_y}{\left(1+\kappa^2\right)^2} & \frac{\kappa(T_y - T_x)}{\left(1+\kappa^2\right)^2}+\frac{\kappa(T_x+T_y)}{2(1+\kappa^2)} \\
\frac{\kappa(T_y - T_x)}{\left(1+\kappa^2\right)^2}- \frac{\kappa(T_x+T_y)}{2(1+\kappa^2)} & \frac{T_y + \kappa^2 T_x}{\left(1+\kappa^2\right)^2} \\
\end{array}\right), 
\label{diffusionmatrix}
\end{equation}
where $\gamma$ is the friction coefficient and $\kappa=qB/\gamma$ is the diffusive Hall parameters quantifying the strength of Lorentz force relative to the frictional force. We consider a reference system at temperature $T$. The two \amnt{effective} temperatures along Cartesian coordinates are measured in units of $T$. The length and time are measured in units of $\sqrt{T/k}$ and $\gamma/k$, respectively.

The applied magnetic field gives rise to velocity correlations which surprisingly remains in the overdamped limit precluding a purely diffusive description of the dynamics as can be realized from the unusual structure of the matrix $\D$ above. Despite the motion of the particle is correlated via the applied magnetic field, there are no stationary-state fluxes in the system. Notice the reduction of the diffusion by a factor $1/(1+\kappa^2)$ for a Brownian system with a single thermostat. This a key feature of Lorentz forces.
\begin{figure}[t]
\centering
\includegraphics[width=1\linewidth]{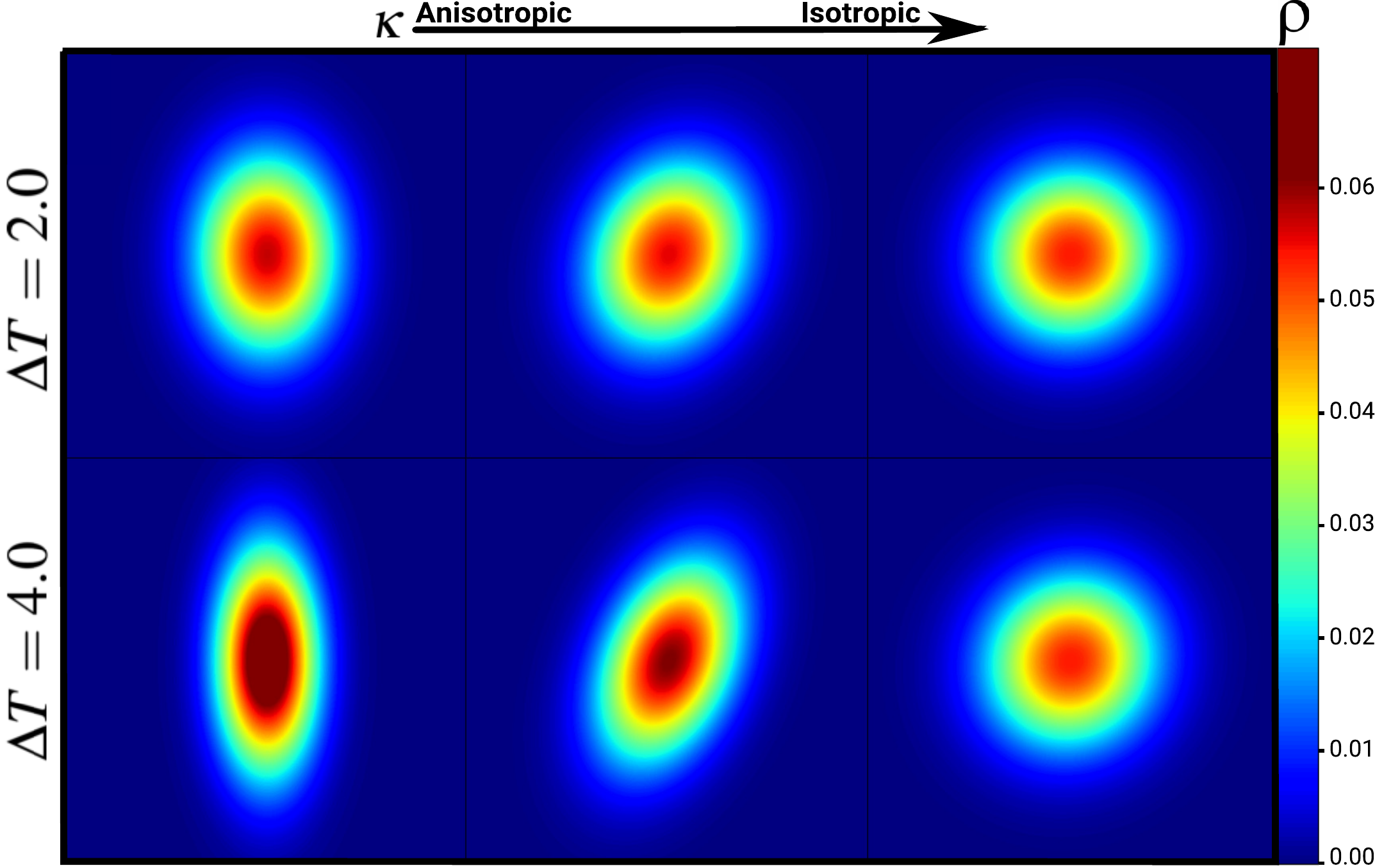}
\caption{The stationary-state probability distribution of the particle's position for two systems with the same \amnt{effective} temperature average $\bar{T}=3.0$ and \amnt{effective} temperature difference $\Delta T=2.0$ for the top panel and $\Delta T=4.0$ for the bottom. 
The broken spatial isotropy in the system by the \amnt{two noises with different strengths} acting on the two spatial particle coordinates is restored in the limit of large magnetic fields. This is attributed to the Lorentz force induced coupling between the spatial degrees of freedom which
makes the \amnt{effective} temperature difference irrelevant at high magnetic fields.  
In each panel, the magnetic field increases from left to right. The diffusive Hall parameter in the left, middle, and right figures takes the values $0.5, 2.0, 10.0$, respectively. In spite of two different \amnt{effective} temperature differences, the two systems settle to the same stationary state where the \amnt{effective} temperature difference becomes irrelavant, as can be seen from the right figures.}
%a charged colloidal particle in an elliptical potential and coupling to two different thermostatsalong perpendicular directions
\label{fig1:pdf}
\end{figure}

%Throughout this work we set the Boltzmann constant $k_B$ to unity. 
Equation~\eqref{FPE} is a linear multivariate Fokker-Planck equation and its stationary-state solution is a Gaussian distribution, which is given as
\begin{equation}
\label{PDF}
\rho_{ss}(\rr) = \frac{1}{Z}e^{-(\rr^\top\cdot\M^{-1}\cdot\rr)},
\end{equation}
where $\M$ is the stationary-state covariance matrix, which can be obtained using Lyapunov approach~\cite{abdoli2020correlations}, and $Z$ is the normalization factor implying the probability density function with total probability of one, given as
\begin{equation}
\label{normalization}
Z=\frac{\pi\sqrt{\kappa^2(T_x+T_y)^2+4T_xT_y}}{k\sqrt{(1+\kappa^2)}}, \hspace{2mm} \text{and} \hspace{2mm}
\M^{-1}=\left( {\begin{array}{*{20}c}
   \mu_1 & \mu_3  \\
   \mu_3 & \mu_2  \\    
 \end{array} } \right),
\end{equation}
where 
\begin{eqnarray}
     \label{mu1}
	\mu_1 &=& k\frac{2T_y+\kappa^2(T_x+T_y)}{\kappa^2(T_x+T_y)^2+4T_xT_y},\\\label{mu2}	
	\mu_2 &=& k\frac{2T_x+\kappa^2(T_x+T_y)}{\kappa^2(T_x+T_y)^2+4T_xT_y},\\
	\label{mu3}
	\mu_3 &=& \kappa (\mu_2-\mu_1).%k\frac{\kappa(T_y-T_x)}{\kappa^2(T_x+T_y)^2+4T_xT_y}.
\end{eqnarray} 

In Fig.~\ref{fig1:pdf}, we show the stationary-state probability distribution of the particle's position from Eq.~\eqref{PDF} for two systems with the same \amnt{effective} temperature average and different \amnt{effective} temperature differences.  In both top and bottom panels, the magnetic field increases from left to right. The stationary states of the systems are flux-free, characterized by a non-Boltzmann probability distribution. The two \amnt{noises with different strengths} acting on the two spatial particle coordinates break the spatial isotropy in the system which is restored in the limit of large magnetic fields. This is attributed to the Lorentz force induced coupling between the spatial degrees of freedom which makes the \amnt{effective} temperature difference irrelevant at high magnetic fields.  
In spite of two different \amnt{effective} temperature differences, the two systems settle to the same stationary state, governed by the average \amnt{effective} temperature, as can be seen from the right figures.

\section{Mean Escape Time}
In this section, we calculate the mean escape time from the two-dimensional truncated harmonic potential at $r=a$ where $r$ is the distance from the origin.
The stationary-state probability density in Eq.~\eqref{PDF} can be represented in polar coordinates as
\begin{equation}
\label{PDFpolar}
\rho_{ss}(r, \theta)= C e^{-\frac{r^2}{2}(\mu_{1} + \mu_{2})}e^{\sqrt{\alpha^2+\beta^2}\sin(2\theta+\phi)},
\end{equation}
where $\alpha=a^2(\mu_2-\mu_1)/2$,  $\beta=\kappa a^2(\mu_2-\mu_1)/2$ and
 $\rho(r, \theta)$ is the probability for the particle to be at $(r, \theta)$.
For the stationary-state probability density $\rho_{ss}(a, \theta)$ on the boundary Eq.~\eqref{PDFpolar} can be rewritten as

\begin{equation}
\label{PDFpolar2}
\rho_{ss}(a, \theta)=C\exp\left[-\frac{ka^2}{\bar{T}(1-\delta_{\kappa}^2)}\big(1+\delta_{\kappa}\sin\left(2\theta+\phi\right)\big)\right], 
\end{equation}
%\cmnt{Derivation of the probability density, role of magnetic field, ...}
where $C=k/(2\pi\bar{T}\sqrt{1-\delta_\kappa^2})$, $\phi=\tan^{-1}(-1/\kappa)$, and $\bar{T}=(T_x+T_y)/2$ is the \amnt{effective} temperature average. Here $\delta_{\kappa}=\Delta T/(2\bar{T}\sqrt{1+\kappa^2})$ is the scaled \amnt{effective} temperature difference, $\Delta T=T_y-T_x$.

In the limit of a large barrier height, the outgoing fluxes is so small that the system settles to a quasistationary state. The quasistationary probability density is given by 
\begin{equation}
\label{quasiss}
P(\rr, t)\sim\rho_{ss}(\rr)e^{-t/\langle t_{esc}\rangle}.
\end{equation}
The outgoing flux can be written as 
\begin{equation}
\label{outgoingflux1}
J(t) \sim V_F a\int_0^{2\pi}\rho_{ss}(a, \theta)e^{-t/\langle t_{esc}\rangle}\dif\theta,
\end{equation}
where $V_F$ is the velocity of the fluctuation path leading to the boundary~\cite{risken1996fokker,woillez2020active}.

Alternatively it is given by
\begin{equation}
\label{outgoingflux1}
J(t) = -\frac{\dif}{\dif t}\int_0^{2\pi}\int_{0}^a P(r, \theta, t)r\dif r\dif\theta,
\end{equation}
where $P(r, \theta, t)$ is the quasistationary probability distribution in the polar coordinates. Using the two equivalent definitions of the outgoing flux the mean escape time is asymptotically equal to the inverse of the stationary-state probability density on the boundary:
\begin{equation}
\label{METdefinition}
\langle t_{esc}\rangle^{-1} \sim V_Fa\int_0^{2\pi} \rho_{ss}(a, \theta)\dif\theta.
\end{equation}

By substituting Eq.~\eqref{PDFpolar2} in Eq.~\eqref{METdefinition} the mean escape time reads
%\begin{equation}
%\label{MET2}
%\
%\end{equation}
\begin{equation}
\label{MET}
\langle t_{esc}\rangle^{-1} \approx \frac{2}{1+\kappa^2}\sqrt{\frac{T\delta_E^2}{\bar{T}(1-\delta_{\kappa}^2)}}e^{-\frac{\delta_E}{1-\delta_{\kappa}^2}}\I_0\left(-\frac{\delta_{\kappa}\delta_E}{1-\delta_{\kappa}^2}\right),
\end{equation}
where $\I_0(x)$ is the modified Bessel function of the first kind of the order zero, $\delta_E=\Delta E/\bar{T}$ is the scaled barrier height, where $\Delta E=ka^2/2$  is the barrier height. Note that the prefactor $\sqrt{2\delta_E}/(1+\kappa^2)$ comes from the fact for a system with the temperature $\bar{T}$ the result reduces to  the Kramers result with the trivial rescaling of the diffusion coefficient, as we show below.  
This can be shown by taking the correct overdamped Langevin equations, derived by taking the small-mass approach, in the presence of an external magnetic field, which is given as~\cite{abdoli2020correlations, chun2018emergence}
\begin{equation}
\label{OverdampedLE}
\dot{\rr}(t) =-k \G^{-1}\rr(t) + \eeta(t),
\end{equation}
where 
\begin{equation}
\label{methods:matrixF}
\G = \gamma\left( \begin{array}{cc}
1 & -\kappa \\
\kappa & 1 \\
\end{array}\right),
\end{equation}
and $\eeta(t)$ is Gaussian nonwhite noise with
\begin{align}
\langle\eeta(t)\rangle & = 0,\\
\langle\eeta(t)\eeta^\top(t')\rangle & = T\G^{-1}\delta_+(t-t') + T(\G^{-1})^\top\delta_-(t-t'),
\end{align}
where the notations $\delta_{\pm}(u=t-t')$ are the modified Dirac delta functions which are zero for $u\neq 0$ while $\int_0^\infty \dif u\delta_+(u)=\int_{-\infty}^0\dif u\delta_-(u)=1$ and $\int_0^\infty \dif u\delta_-(u)=\int_{-\infty}^0\dif u\delta_+(u)=0$. 

Using the Ito calculus the Langevin equations in Eq.\eqref{OverdampedLE} are reduced to the one-dimensional problem for the variable $r=|\rr|$, which is given as~\cite{gardiner2009stochastic}

\begin{equation}
\label{1DLE}
\dif r =\left(-\frac{k}{\gamma(1+\kappa^2)}r+\frac{\mathrm{D}_0}{1+\kappa^2}\frac{1}{r}\right)\dif t+\sqrt{\frac{2\mathrm{D}_0}{1+\kappa^2}}\xi(t)\dif t,
\end{equation}
where $\mathrm{D}_0=T/\gamma$ is the coefficient of a freely diffusing particle. Note that the stochastic force in Eq.~\eqref{1DLE} is Gaussian white noise. 
The mean escape time for the one-dimensional stochastic differential equation Eq.~\eqref{1DLE} can be written as 
\begin{equation}
\label{MFPTintegral}
\langle t_{esc}\rangle_{eq} =\frac{2(1+\kappa^2)}{\mathrm{D}_0}\int_0^a y^{-1}e^{\frac{ky^2}{2\gamma\mathrm{D}_0}}\dif y\int_0^y ze^{-\frac{kz^2}{2\gamma\mathrm{D}_0}}\dif z,
\end{equation}
where the subscript \textit{eq} shows the mean escape time due to the equilibrium fluctuations. This can be exactly solved and the solution reads
\begin{equation}
\label{MFPTeqm}
\langle t_{esc}\rangle_{eq} = \frac{\gamma(1+\kappa^2)}{2k}\left[\Ei\left(\frac{\Delta E}{\gamma\mathrm{D}_0}\right)-\log\left(\frac{\Delta E}{\gamma\mathrm{D}_0}\right)-\gamma_{EM}\right]
\end{equation} 
where $\gamma_{EM}$ is the Euler-Mascheroni constant \amnt{and $\Ei(x)$ is the exponential integral. The series expansion of the exponential integral at $x=\infty$ is $e^x (1/x + (1/x)^2 + 2/x^3 + 6/x^4 + 24/x^5 + O((1/x)^6))$. Therefore, in the limit of large barrier heights the exponential integral in Eq.~\eqref{MFPTeqm} can be written as $\Ei\left(\Delta E/\gamma \mathrm{D}_0\right)\sim \exp(\delta_E)/\delta_E$. As a consequence, Eq.~\eqref{MFPTeqm} reduces to  $\sim(1+\kappa^2)\gamma e^{\delta_E}/2k\delta_E$ in the limit of large barrier heights. The large barrier height approximation is highly accurate when the expression is scaled by an empirical factor $1.2$. }

\begin{figure}[t]
\centering
\includegraphics[width=1\linewidth]{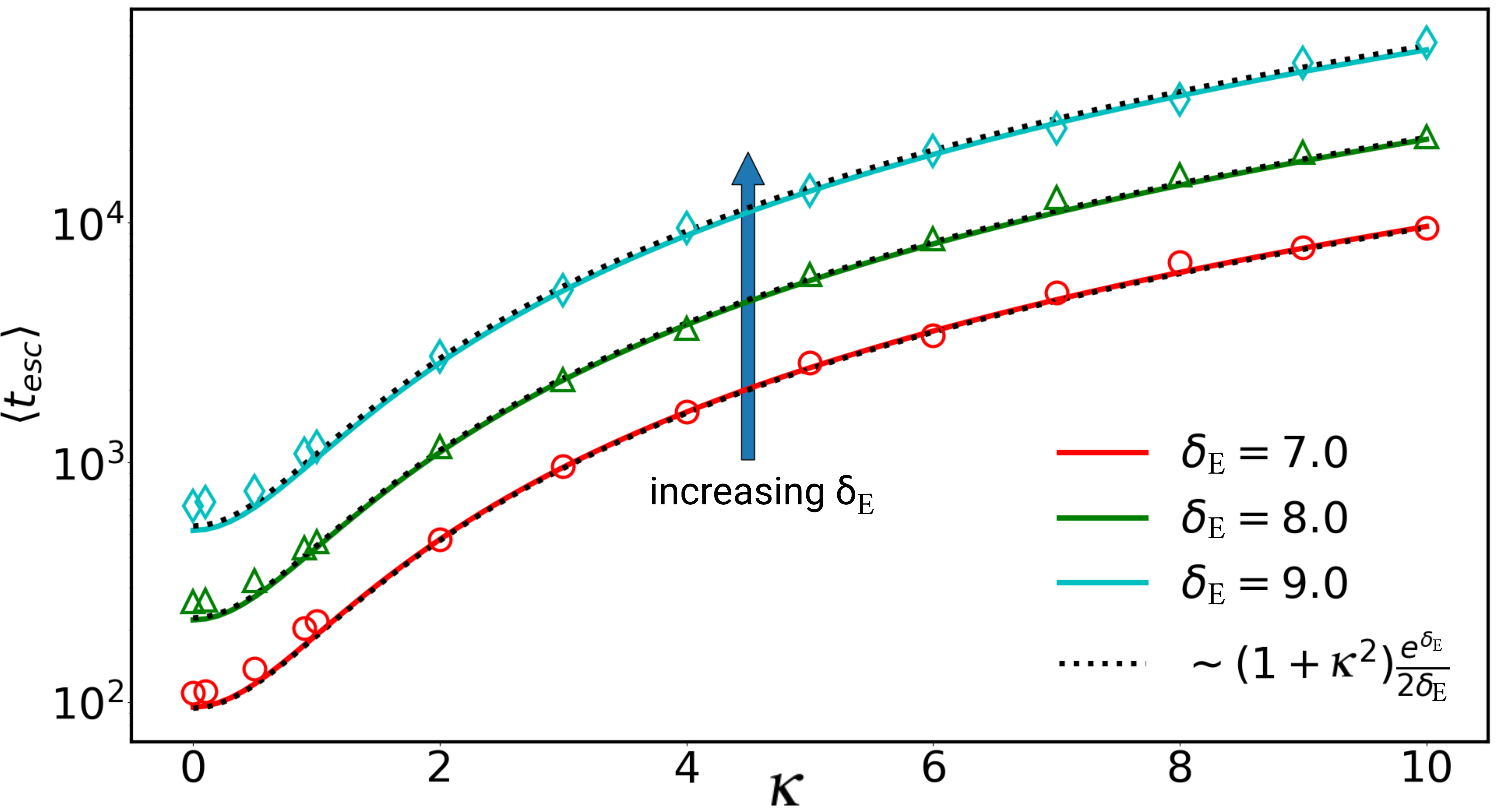}
\caption{The mean escape time as a function of the diffusive Hall parameter $\kappa$ for different values of the scaled barrier heights $\delta_E$ in a system at the temperature $T=3.0$. The solid lines show the theoretical prediction from Eq.~\eqref{MFPTeqm} and the symbols depict the simulation results. The dotted lines show the approximate result in the limit of large barrier heights (i.e., Kramers results) which is in a good agreement with the exact solution. Note that the temperature difference vanishes here.
}
%a charged colloidal particle in an elliptical potential and coupling to two different thermostatsalong perpendicular directions
\label{fig2}
\end{figure}

Figure~\ref{fig2} shows the mean escape time of a system with temperature $T=3.0$ as a function of the diffusive Hall parameter $\kappa$. As can be seen, the theory (shown by solid lines) matches with the simulation results (shown by symbols). The large barrier approximation is accurate when the expression is scaled by $1.2$. We use the same prefactor for the asymptotic approach used in this paper.

%\subsection{Dimensions} We consider a reference system at temperature $T$ such that the temperatures along the perpendicular spatial directions are measured in the units of $T$. The dimension of the probability density  in Eq.~\eqref{METdefinition} is the inverse of area which is evident from the factor $C$ in Eq.~\eqref{PDFpolar2}, which has the dimension of $\left[k/\bar{T}\right]=1/a_\circ^2$. The unit of length $a_0$, can be measured from the fact that $ka_\circ^2$ has the unit of $T$ such that $a_\circ=\sqrt{T/k}$. Thus $a$ can be defined as $a_\circ\breve{a}$ where $\breve{a}=\sqrt{2\Delta E/T}$. The unit of velocity is $a_\circ/t_\circ$ where $t_\circ$ is given in the unit of $\gamma/k$. Taking into account all these units, one gets the mean escape time in $t_\circ$.
%\acknowledgments
%Insert here the text.
%\bibliographystyle{eplbib}
%\bibliography{references}
\providecommand{\noopsort}[1]{}\providecommand{\singleletter}[1]{#1}%